\documentclass[letterpaper, 12 pt]{article}
\usepackage[a4paper, total={6in, 8in}]{geometry}
\usepackage{graphicx} 
\usepackage{amsmath}
\usepackage{graphicx}
\usepackage{cancel}
\usepackage{lipsum}
\usepackage{cite}
\usepackage[dvipsnames]{xcolor}
\usepackage{fancyhdr}

\title{Tunneling of Hawking Radiation in Starobinsky-Bel-Robinson gravity}
\author{Dhanvarsh Annamalai\footnote{dhanvarshannamalai2002@gmail.com} and Akshat Pandey\footnote{apandey.physics@gmail.com}  \\ \normalsize Department of Physics, Shiv Nadar Institution of Eminence \\ \normalsize Greater Noida, Uttar Pradesh-201314, India} 
\date{\today\footnote{This is a preprint of the work accepted for publication in Gravitation and Cosmology, © copyright (2024), the copyright holder indicated in the Journal.}}

\begin{document}

\maketitle

\begin{abstract}
We examine Hawking radiation for a Schwarszchild-type black hole in Starobinsky Bel Robinson (SBR) gravity and calculate the corrected Hawking Temperature using the tunnelling method. We then discuss the deviation of our Hawking temperature from the standard Schwarszchild result. We relate the corrections to the Hawking temperature beyond the semi-classical approximation. We highlight that starting with a modification of the classical black hole geometry and calculating the semi-classical Hawking temperature, yields temperature corrections comparable to those obtained when the classical background is kept unchanged and beyond semi-classical terms in the temperature are included.

\end{abstract}

\section{Introduction}

Ever since Hawking's initial work \cite{Hawking, Hawking temp derivation}, the physics of black hole radiance has seen continued interest. In their seminal paper, Parikh and Wilzeck proposed a heuristic derivation for the Hawking temperature using the WKB approximation where Hawking radiation was thus described as tunnelling \cite{Parikh}. This helped emphasise the simplicity of the underlying physics of Hawking radiation. For the application of this method to derive Hawking temperature in various black hole solutions, see \cite{Parikh 2, Feng 1, Feng 2, Feng 3, Flan, Bagchi, Wang, Usage of tunnelling}. 

A key feature of the above mentioned tunnelling method is that it is performed within a semi-classical scenario. Indeed, these results have later been generalised beyond the semi-classical approximation, thus including higher-order corrections. For instance, Banerjee and Majhi \cite{Manji 2}  obtained the higher order corrections to the Hawking temperature of black hole solutions within GR using the Hamilton-Jacobi method. 
In particular, for the Schwarszchild black hole, written in gravitational units, the corrected Hawking temperature $T_H$ has the form,

\begin{equation}
    T_H = \frac{1}{8 \pi G M} \left(1 + \sum_{i} \frac{\lambda_i}{G^iM^{2i}} \right)^{-1}
\end{equation}

Here the first term on the right-hand side is the one obtained via semi-classical approximation. The subsequent terms arise as quantum corrections, the first among which corresponds to the effects of backreaction \cite{Manji 1}. For more details on this method see \cite{Manji 2, Modak, Manji 3}

In the present work, we try to explore the other path, that is to answer the following question \textbf{---} if we start with a modified black hole solution, which includes quantum gravity-inspired corrections in the metric, and use simply the semi-classical approximation, how close can we come to reproducing the higher order correction terms in the Hawking temperature?

For this purpose, we turn towards Starobinsky-Bel-Robinson (SBR) gravity \cite{SBR}, which has been gaining interest recently. SBR gravity is a quantum-gravity-inspired modification to the usual Einstein-Hilbert (EH) action. As the name suggests, this includes the addition of an $R^2$ term like the one in Starobinsky's inflationary model \cite{Star}, and another term related to the Bel-Robinson tensor from M-theory \cite{Bel}.

More recently, Ketov et al \cite{SBR Blackhole} showed that SBR gravity allows for the presence of a modified Schwarszchild-like spherically symmetric black hole solution. The modifications to the black hole metric are a consequence of the modifications to the EH action as mentioned above. This is exactly the type of solution we need, to answer the question that we have asked above. In the following sections, we will make use of this black hole solution to calculate the Hawking temperature of this black hole solution via the Parikh-Wilzceck method and compare it against the results available in the literature.


This paper is organised as follows. In section 2, we briefly review SBR gravity and the Schwarzschild-like black hole solution as obtained by Ketov et al \cite{SBR Blackhole}. In section 3, we employ the semi-classical tunnelling method to calculate the transmission rate for Hawking radiation. In section 4, we discuss the relevant thermodynamical quantities and their associated corrections. We compare our results against the expected corrections from \cite{SBR Blackhole}. We also relate our results to Banerjee et al \cite{Manji 2} as mentioned above. We end with a summary of our work in section 5. 

\section{SBR gravity and a black hole solution}

The SBR action reads

\begin{equation}
    S_{SBR}= \frac{M_{\mathrm{Pl}}^2}{2} \int d^4 x \sqrt{-g}\left[R+\frac{1}{6 m^2} R^2-\frac{\beta}{32 M_{\mathrm{Pl}}^6}\left(\mathcal{P}^2-\mathcal{G}^2\right)\right]
\end{equation}

Here, $\beta > 0$ is the new dimensionless coupling constant to be determined by compactification of M-theory, $m$ is the inflation mass  $\sim 10^{-5} M_{pl}$ determined by the COBE/WMAP normalization, where the reduced Planck mass $M_{\mathrm{Pl}}=1/\sqrt{8 \pi G}$.The $R^2$ term corresponds to the Starobinsky modification to the Hilbert action \cite{Star}. $\mathcal{P}$ and $ \mathcal{G}$ are the Euler and Pontryagin topological densities in \textit{D} = $4$ dimensions, respectively such that 
\begin{equation}
    \mathcal{G}=R^2-4 R_{\mu \nu} R^{\mu \nu}+R_{\mu \nu \rho \sigma} R^{\mu \nu \rho \sigma}
 \end{equation} 
 and 
 \begin{equation}
 \mathcal{P}=\frac{1}{2} \sqrt{-g} \epsilon_{\mu \nu \rho \sigma} R_{\alpha \beta}^{\rho \sigma} R^{\mu \nu \alpha \beta} .
\end{equation} 

These are related to the squared Bel-Robinson tensor $\mathcal{T}^{\alpha \beta \mu \nu }$ \cite{Bel, Robinson}  through
\begin{equation}
    \mathcal{T}^2 = \frac{1}{4}(\mathcal{P}^2 - \mathcal{G}^2)
\end{equation}

For details on equation (5) and its significance, see \cite{deser}.

The purely gravitational SBR action (2) has several interesting features including its robustness and predictive power given the presence of only two parameters $m$ and $\beta$. However, in the present work, we are interested in the corrections this action imposes on the Schwarszchild black hole metric.   

The $\beta$-corrected Schwarzschild metric as worked out by Ketov et al \cite{SBR Blackhole} is given by
\begin{equation}
    ds^2= -\mathcal{F}(r)dt^2 + \frac{1}{\mathcal{F}(r)}dr^2 + r^2 d\Omega ^2
\end{equation}
Where $\mathcal{F}(r)$ corresponds to 
\begin{equation}
   \mathcal{F}(r)= 1 - \frac{r_s}{r} + \beta \frac{128 \pi ^3}{5}\left(\frac{G r_s}{r^3} \right)^3\left(108 - 97\frac{r_s}{r} \right)
\end{equation}

Here $r_s$ is the standard Schwarzschild radius from GR, $r_s=2GM$. 

Given the $\beta$ correction to the metric components, it is natural to expect that the black hole corresponding to it will also have corrections to its physical properties, like the radius, temperature, entropy, etc. As we shall see in the next section, it is essentially the form of $\mathcal{F}(r)$ that governs the imaginary action and consequently the tunneling rate; a modification to $\mathcal{F}(r)$ would thus imply modifications to the tunneling rate.

As \cite{SBR Blackhole} showed, the black hole radius $r_H$ with corrections up to first order in $\beta$ is

 \begin{equation}
     r_H = 2GM - \beta \frac{44 \pi^3 }{5G^2 M^5} + \textit{O}(\beta^2)
 \end{equation}

Therefore, there is a shift in the horizon given by the $\beta$ term. We are interested in the corrections induced because of the $\beta$ term upon the Hawking temperature. As we will see in the next section, up to the first order in $\beta$, the only relevant correction is the one associated with the radius. The problem we address, thus, reduces to the following \textbf{---} given that the radius of a Schwarszchild-like black hole is prescribed by equation (7), how does this alter the Tunneling coefficient and in turn the Hawking temperature?

\section{The Tunneling coefficient}

We begin by transforming equation (6) to Painlev$e^{'}$ coordinates. This, among other things, eliminates the coordinate singularity at the horizon. 

\begin{equation}
    ds^{2} = - \mathcal{F}(r) dt^{2}_{p} + 2 \sqrt{(1-\mathcal{F}(r))} dt_{p} dr+ dr^2 + r^2 d\Omega^2
\end{equation}
In (9), $t_{p}$ denotes the Painlev$e^{'}$ time coordinate. Then the radial null geodesic is given by,

\begin{equation}
    ds^2 = 0 = - \mathcal{F}(r) dt^{2}_{p} + 2 \sqrt{(1-\mathcal{F}(r))} dt_{p} dr + dr^2 
\end{equation}
hence we have,
\begin{equation}
    - \mathcal{F}(r) + 2 \sqrt{(1-\mathcal{F}(r))}  \dot{r} + \dot{r}^2 = 0
\end{equation}
thus,
\begin{equation}
    \dot{r} = \pm 1 - \sqrt{(1-\mathcal{F}(r))}
\end{equation}
 The positive (negative) sign refers to outgoing (incoming) geodesics. The imaginary part of the action over the classically forbidden path is given by, 
\begin{equation}
    \text{Im}S = \text{Im} \int^{r_{out}}_{r_{in}} p \, dr 
\end{equation}
\begin{equation}
    \text{Im}S = \text{Im} \int^{r_{out}}_{r_{in}}  dr \int^{p}_{0} dp
\end{equation}
Here $r_{in}$ is slightly inside the black hole, where the particle is produced and $r_{out}$ is slightly outside the shrunken radius of the black hole after the particle has tunnelled through it. Changing the order of integration and the variables from momentum to energy via Hamilton's equation we get, 
\begin{equation}
    \text{Im}S = \text{Im}  \int^{M-\omega}_{M}  dH \int^{r_{out}}_{r_{in}} \frac{1}{\dot{r}} dr = \text{Im}   - \int^{\omega}_{0} d\omega^{'} \int^{r_{out}}_{r_{in}} \frac{1}{(1-\sqrt{(1-\mathcal{F}(r))}} dr
\end{equation}
where we have used equation (11). Now, since the radius is defined by $\mathcal{F}(r) = 0$, and we care only about the near horizon behaviour i.e. $\mathcal{F}(r) << 1$, we can safely use the following approximation  $ \frac{1}{(1-\sqrt{(1-\mathcal{F}(r))}} \approx \frac{2}{\mathcal{F}(r)}$. Thus, using equation (7) and Taylor expanding around $r_H$ up to the first order, $\mathcal{F}(r)$ turns out to be,
\begin{equation}
    \mathcal{F}(r) \approx  \left[r - 2G(M-\omega^{'}) - \beta \frac{44 \pi^3 }{5G^2 (M-\omega^{'})^5}\right] \, \partial_{r}\mathcal{F}(r) 
\end{equation}
The derivative is taken at $r=r_{H}$ which up to linear order in $\beta$ is given by 
\begin{equation}
    \partial_{r}\mathcal{F}|_{r = r_{h}} = \frac{1}{2G(M-\omega^{'})} + \frac{\beta \pi^2}{G^4 (M-\omega^{'})^{7}} 
\end{equation}
Note that the integration limits are 
\begin{equation}
    r_{in} = 2GM - \beta \frac{44 \pi^3 }{5G^2 M^5} - \epsilon \, , r_{out} =  2G(M-\omega^{'}) - \beta \frac{44 \pi^3 }{5G^2 (M-\omega^{'})^5} + \epsilon 
\end{equation}
Check ref [23] for further details on this change in the black hole radius, in the presence of a scalar field. Substituting (16) in (15) and changing the integration variable to $\epsilon$,

\begin{equation}
    \text{Im}S = \text{Im} \int^{\omega}_{0} -d\omega^{'}  \int^{0^{+}}_{0^{-}} \frac{2}{\epsilon\partial_{r} \mathcal{F}|_{r = r_{h}}}  d\epsilon = \int^{\omega}_{0} d\omega^{'}   \frac{4\pi G(M-\omega^{'})}{ \left(1 + \frac{2 \beta \pi^2 }{G^3 (M-\omega^{'})^{6}}\right) }  
\end{equation}
Where we have used the property that $\text{Im}(\frac{1}{\epsilon}) = i\pi \delta(\epsilon)$ for this integral. Upon Taylor expansions and keeping terms only up to linear order in $\beta$, we get
\begin{equation}
    \text{Im}S \approx  \int^{\omega}_{0} d\omega^{'} 4\pi G (M-\omega^{'}) - \beta \int^{\omega}_{0} d\omega^{'} \frac{8\pi^3}{G^2 (M-\omega^{'})^5}
\end{equation} 
The imaginary part of the action, thus, comes out to be
\begin{equation}
    \text{Im}S = 4\pi G \omega\left(M - \frac{\omega}{2}\right) -  \frac{2 \pi^3 \beta }{G^2} \left[  \frac{1}{ (M-\omega)^4} - \frac{1}{M^4} \right]
\end{equation}
using this, the exponential part of the semi-classical tunnelling rate
\begin{equation}
    \Gamma \sim exp(-2 \text{Im}S) \sim exp\left( -8\pi G \omega\left(M - \frac{\omega}{2}\right) + \frac{4 \pi^3 \beta }{G^2} \left[ \frac{1}{ (M-\omega)^4} - \frac{1}{M^4} \right] \right)
\end{equation}
neglecting the quadratic and higher order terms in $\omega$, we get
\begin{equation}
    \Gamma  \sim exp\left( -8\pi G \omega M + 16 \pi^3 \beta  \frac{\omega}{G^2 M^5} + \omega \textit{O}(\beta^2)\right)
\end{equation}
This is our desired Boltzmann factor for a particle with energy $\omega$. In the next section, we will use this exponential to define the Hawking temperature in terms of the coefficients of $\omega$.

\section{Hawking temperature}

Upon identifying (23) with the definition of the Boltzmann factor $exp(-\omega / T_H)$, we get the Hawking temperature to be

\begin{equation}
   {T}_{H} = \frac{1}{8\pi G M} + \frac{\beta \pi} {4 G^4 M^7} + \textit{O}(\beta^2)
\end{equation}
Some remarks are in order:

\begin{itemize}
    \item There are corrections to the semi-classical Hawking temperature induced by the Bel-Robinson coupling constant $\beta$. This was to be expected and somewhat trivial.
    \item In their paper, Delgado and Ketov [2] calculated the Hawking temperature in terms of the surface gravity, using the derivative of the metric coefficient $\frac{d \mathcal{F}(r) }{dr}$. Comparing this result with what we have obtained above, we see that we get the correct mass dependence $(M^{-7})$ in the $\beta$ term. However, we are off by a factor of $4 \pi$ in the coefficients of this term. This is a consequence of the crudeness of the semi-classical approximation which we have employed.
    \item Comparing equation (24) with equation (1), we see that the $\beta$ correction term to the Hawking temperature can be interpreted in terms of Hamilton-Jacobi-based quantum corrections to the semi-classical Hawking temperature, upon making the identification 
    \begin{equation}
        \beta \sim - \lambda_3
    \end{equation}
    while ignoring all the other $\lambda_i$ terms.
    Thus, we see that the same modification of Hawking temperature can be interpreted as either due to a modification of the black hole geometry or due to the modification of the semi-classical structure of the tunnelling approximation [3].  
    
\end{itemize}

\section{Summary}

In this work, we employed the semi-classical method to calculate the tunnelling coefficient for a Schwarszchild-like black hole in SBR gravity. This allowed us to calculate the Hawking temperature and its deviations from the standard semi-classical result. Further, we compared our corrections results with those existing in the literature. We emphasize that commencing with a modification of the classical black hole geometry and incorporating the semi-classical Hawking temperature,  can yield corrections to the Hawking temperature which is analogous to maintaining the classical background unaltered and, instead, going beyond the semi-classical approximation itself. These corrections are of course, similar up to dimensionless parameters.

\section*{Acknowledgements}
We would like to express our sincere gratitude to Kaustubh Singh and Sauvik Sen for their insightful comments and valuable feedback. We also thank the anonymous reviewers for their constructive criticism and helpful suggestions.


\begin{thebibliography}{20}
\small

\bibitem{Hawking} 
S. W. Hawking, 
"Particle creation by black holes",
\textit{Comm. Math. Phys.}, \textbf{43}, 199 (1975).

\bibitem{Hawking temp derivation}
 J. B. Hartle, \&  S. W. Hawking, "Path-integral derivation of black-hole radiance", \textit{Phys. Rev. D} \textbf{13}, 2188 (1976).

\bibitem{Parikh} 
M. K. Parikh,  \&  F. Wilczek, "Hawking Radiation As Tunneling", \textit{Phys. Rev. Lett.}, \textbf{85}, 5042 (2000).

\bibitem{SBR Blackhole} 
 R. C. Delgado, \&  S. Ketov, "Schwarzschild-type black holes in Starobinsky-Bel-Robinson gravity", \textit{Phys. Lett. B.}, \textbf{838}, 137690 (2023).

\bibitem{Manji 1} 
 R. Banerjee, \&  B. R. Majhi, "Quantum tunneling beyond semiclassical approximation", \textit{J. High Energy Phys.}, \textbf{2008}(06) (2008).

\bibitem{Manji 2} 
 R. Banerjee, \&  B. R. Majhi, "Quantum tunneling and back reaction", \textit{Physics Letters B.}, \textbf{662}, 0730-2693 (2008).

\bibitem{Manji 3} 
 B. R. Majhi, "Fermion tunneling beyond semiclassical approximation"  \textit{Phys. Rev. D}, \textbf{79}(4), 044005 (2009).

\bibitem{Modak} 
 S. K. Modak, "Corrected entropy of BTZ black hole in tunneling approach",  \textit{Phys. Lett. B.}, \textbf{671}(1), 167-173 (2009).

\bibitem{Parikh 2} 
 M. Parikh, "A Secret Tunnel Through the Horizon", \textit{General Relativity and Gravitation} \textbf{36}, 2419-2422 (2004).

\bibitem{Feng 1} 
 Z. W. Feng,  H. L. Li,  X. T. Zu, \&  S. Z. Yang, "Quantum corrections to the thermodynamics of Schwarzschild–Tangherlini black hole and the generalized uncertainty principle", \textit{European Physical Journal.}, \textbf{C76}, 212 (2016).

\bibitem{Feng 2} 
 Z. W. Feng,  Q. C. Ding, \&  S. Z. Yang, 
\textit{European Physical Journal.}, \textbf{C79}, 445 (2019).

\bibitem{Feng 3} 
 Z. W. Feng,  X. Zhou,  S. Q. Zhou, \&  D. D. Feng, "Modified fermion tunneling from higher-dimensional charged AdS black hole in massive gravity", \textit{Annals of Physics}, \textbf{416}, 168144 (2020).

\bibitem{Flan} 
 E. E. Flanagan, "Order-Unity Correction to Hawking Radiation", \textit{Phys. Rev. Lett.}, \textbf{127}, 041301 (2021).

\bibitem{Bagchi} 
 B. Bagchi, \&  S. Sen, "Tunneling of hawking radiation for BTZ black hole revisited", \textit{Int. J. Mod. Phys.}, \textbf{37}(02) (2022).

\bibitem{Wang} 
 R. Li, \&  J. Wang, "Hawking radiation, local temperatures, and nonequilibrium thermodynamics of the black holes with non-Killing horizon",
\textit{Phys. Rev. D.}, \textbf{104} , 026011 (2021).

\bibitem{Usage of tunnelling} 
 Z. Z. Ma, "Hawking temperature of a Kerr–Newman–dS black hole from tunneling",
\textit{Class. Quantum Grav.}, \textbf{26}, 045002 (2009).

\bibitem{Visser} 
M. Visser, "ESSENTIAL AND INESSENTIAL FEATURES OF HAWKING RADIATION"
\textit{Int. J. Mod. Phys. D.}, \textbf{12}(04) (2003).

\bibitem{SBR} 
 S. V. Ketov, "Starobinsky–Bel–Robinson Gravity",
\textit{Universe}, \textbf{8}, 351 (2022).

\bibitem{Star} 
 A. A. Starobinsky,"A new type of isotropic cosmological models without singularity",
\textit{Phys. Lett. B}, \textbf{91}, 99–102  (1980).

\bibitem{Bel}
 L. Bel, 
\textit{Colloq. Int. Cent. Natl. Rech. Sci.}, \textbf{91}, 119-126 (1962).

\bibitem{Robinson} 
 I. Robinson, "On the Bel - Robinson tensor",
\textit{Class. Quantum Gravity}, \textbf{14}, A331–A333 (1997).

\bibitem{deser}  S. Deser, "The Immortal Bel-Robinson Tensor",  arXiv: 9901007.

\bibitem{Nozari} 
Eslamzadeh, Sareh, and Kourosh Nozari, "Tunneling of massless and massive particles from a quantum deformed Schwarzschild black hole surrounded by quintessence", \textit{Nuclear Physics B} {\bf 959:} 115136 (2020).
\end{thebibliography}
\end{document}